# Van den Bergh 24 is a variable reflection nebula around XY Per

Harald Strauß[1], Klaus Bernhard[2]

1) Astronomischer Arbeitskreis Salzkammergut
email: h.strauss@aon.at
2) Bundesdeutsche Arbeitsgemeinschaft für Veränderliche Sterne e.V., Germany
American Association of Variable Star Observers (AAVSO), USA
email: klaus.bernhard@liwest.at , ORCID 0000-0002-0568-0020

**Abstract:** We report the discovery of optical variability in vdB 24 on timescales of a few days. This reflection nebula surrounds the Herbig Ae/Be star XY Per, whose changing brightness appears to influence the nebular illumination.

## 1   Introduction

Reflection nebulae are interstellar clouds of gas and dust that do not shine by their own emission but by reflecting the light of nearby stars [1]. In contrast to emission nebulae, the radiation of the embedded stars is insufficient to ionize the surrounding gas; instead, starlight is scattered by fine dust particles. Typically, reflection nebulae appear bluish, since short-wavelength blue light is scattered more efficiently than longer-wavelength red light. Well-known examples include the bright nebula Messier 78 in Orion and the nebulosity around the Pleiades with relatively bright and persistent cloud structures.

These nebulae exhibit morphological changes only over timescales of hundreds or thousands of years. However, there exists a special class of variable reflection nebulae that show significant changes in brightness and shape on shorter timescales—within years, months, or even weeks. These objects are almost always associated with young variable stars whose instability affects the appearance of the surrounding nebula. Historically, these phenomena have been among the most fascinating in nebular observation. Already in 1783, William Herschel noted a nebulous patch in the constellation Monoceros that would later become known as Hubble's Variable Nebula (NGC 2261, = R Mon) [2]. In the second half of the 19th century, another unusual nebula was discovered: John Russell Hind reported in 1852 on a reflection nebula near the young star T Tauri in Taurus. This Hind's Variable Nebula (NGC 1555) disappeared and reappeared over the decades, suggesting a connection to the variable nature of the embedded star [3]. These early observations laid the foundation for the understanding that some nebulae change their brightness and appearance in a dynamic way driven by processes in and around young stars. In recent times, variable nebulae continue to be discovered, such as the nebula surrounding NGC 1333 in the Perseus region [4]. The observed brightness variations of variable reflection nebulae can be explained by the following, often interrelated, mechanisms:

-Intrinsic variability of the central star: Young stars such as T-Tauri and Herbig Ae/Be stars often show irregular fluctuations in their radiation output.

-Accretion-driven outbursts: Especially strong brightening events occur when a protostar undergoes an accretion burst, as seen in FU Orionis or EX Lupi–type stars. These cause the luminosity of the star to increase by several magnitudes over short timescales. A previously faint reflection nebula may suddenly light up dramatically.

-Dust structures and shadow effects: As in the case of Hubble's Nebula, dense dust clumps or features in the circumstellar disk can act as variable "shutters." As these structures move (due to disk rotation or other dynamics), they cast moving shadows into the nebula. This leads to brightness and structural changes in the nebula even if the star's total brightness remains unchanged – a form of "cosmic shadow play".

We report the discovery of optical variability in vdB 24 on timescales of a few days. To the best of our knowledge, based on a review of the literature via SIMBAD and NASA ADS, this variability has not been reported previously.

## 2 Observation and data analysis

The initial dataset for this study was derived from the CCD-Guide, an initiative of the Astronomical Working Group Salzkammergut[1], comprising approximately 850 entries of known reflection nebulae. These objects were subsequently sorted in a spreadsheet according to custom criteria, with a primary focus on apparent size, as brightness variations are typically confined to regions near the central illuminating star.

Reflection nebulae located in the southern celestial hemisphere were excluded from further consideration, as they are not accessible from the Gahberg observatory in Austria and are not covered by the Palomar Observatory Sky Survey (POSS-I) [2]. This survey, based on photographic plates taken with the 48-inch Oschin Schmidt telescope at Mount Palomar in southern California, only includes the northern sky.

Based on the aforementioned list, images from the NASA STScI Digitized Sky Surveys DSS1 and DSS2, which are based on photographic data obtained using the Oschin Schmidt Telescope on Palomar Mountain and the UK Schmidt Telescope, were retrieved and blinked for comparison. As the images are well-aligned, they can generally be used directly without the need for additional adjustment. This analysis revealed a particularly pronounced apparent variability in the reflection nebula Van den Bergh 24 (=vdB 24) surrounding the variable star XY Per. An animation based on these DSS1 and DSS2 images is accessible alongside the online version of this paper.

To investigate the brightness variations of VdB 24 in detail, images from the Zwicky Transient Facility (ZTF) were used, a time-domain survey that has been operational since 2017 at the Palomar Observatory. The ZTF camera, featuring e2v CCD231-C6 devices, is mounted on the Palomar 48-inch Samuel Oschin Schmidt Telescope. Covering three passbands (g, r, and partially i), it achieves a limiting magnitude of 20.5 mag, making ZTF data highly suitable for variable-objects investigations [5-7].

---

[1] https://ccdguide.com
[2] https://skyserver.sdss.org/dr5/en/proj/advanced/skysurveys/poss.asp

Unfortunately, the central star is already overexposed in the ZTF images, so other surveys—namely KWS (Kamogata/Kiso/Kyoto Wide-field Survey) and ASAS-SN (All Sky Automated Survey for SuperNovae) were used to compare the brightness evolution of the nebula with that of the central star [9-11]. Therefore, the photometric data are not available for exactly the same dates as the ZTF images. However, they are sufficient to compare the brightness evolution of the central star with that of the nebula over a time span of several days and weeks.

XY Per is a visual binary system with a Herbig Ae primary and a B6Ve secondary separated by 1.4 arcsec [12]. The primary star exhibits irregular, eruptive variability characteristic of the Orion type variables. The reflection nebula vdB 24 surrounds XY Per and is illuminated by it. vdB 24 appears to be divided into three reflection nebulae with bluish colour, due to the scattering of starlight by interstellar dust. The part of the nebula identified as variable is the central region directly surrounding the central system XY Per (center of Figure 1). This nebula is embedded within the larger dark nebula LDN 1442 (brownish) and is part of the Perseus molecular cloud complex [13].

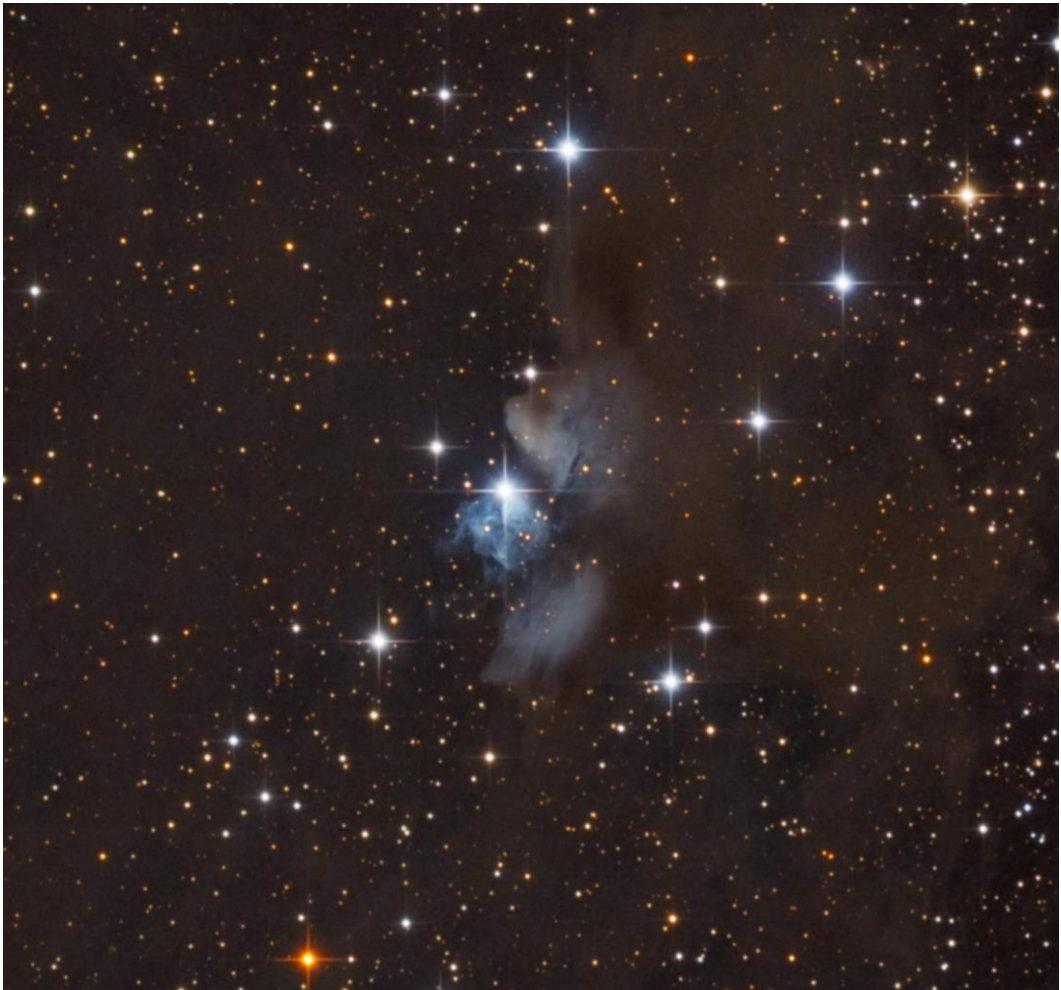

Figure 1: Overview image of the VdB 24 region, Bernhard Hubl | CCDGuide.com
(North is up, 27' x 26')

The essential data of the central binary star system XY Per are listed in Table 1:

Table 1: Essential data of XY Per

| | |
|---|---|
| Spectral type | XY Per A: Herbig AeBe [8,12] |
| | XY Per B: B6Ve [12] |
| Identifiers | Gaia DR3 223935338503901952 (XY Per A) |
| | Gaia DR3 223935338501793664 (XY Per B) |
| | LAMOST DR9 J034936.63+385859.0 |
| | PPM 68809* |
| | HIP 17890* |
| | TYC 2863-2195* |
| | GSC 02863-02195* |
| | ASAS-SN J034936.30+385854.5 |

*PPM: Positions and Proper Motions Catalog; HIP: Hipparcos; TYC: Tycho; GSC: Guide Star Catalog

| | | |
|---|---|---|
| Magnitude Range (V) | 9.2-10.7 (derived from the KWS [9], star A+B combined) | |
| Gaia DR3 [14]: | Right Ascension (J2000) | 03 49 36.33659 (XY Per A)* |
| | Declination (J2000) | +38 58 55.5483 (XY Per A)* |
| | Plx | 2.3537±0.0302 mas (XY Per A) |
| | Gmag | 9.4994 mag (XY Per A) |
| | Right Ascension (J2000) | 03 49 36.44723 (XY Per B)* |
| | Declination (J2000) | +38 58 55.8581 (XY Per B)* |
| | Plx | 2.3623±0.0491 mas (XY Per B) |
| | Gmag | 10.7885 mag (XY Per B) |

* All coordinates are taken from the Gaia DR3 catalogue (http://vizier.u-strasbg.fr/viz-bin/VizieR?-source=I/355). The coordinates (epoch J2000) are computed by VizieR, and are not part of the original data from Gaia (note that the coordinates are computed from the positions and the proper motions).

## 3 Results

The photometric data from the KWS survey in the Ic and V bands between 2011 and 2025 show that the central star XY Per (combined light from companion stars A + B) is consistently irregularly variable to a moderate degree (up to ~0.5 mag), but exhibited a higher amplitude of up to 1.5 mag within a few weeks during the years 2011–2013 and 2020–2022 (Figure 2).

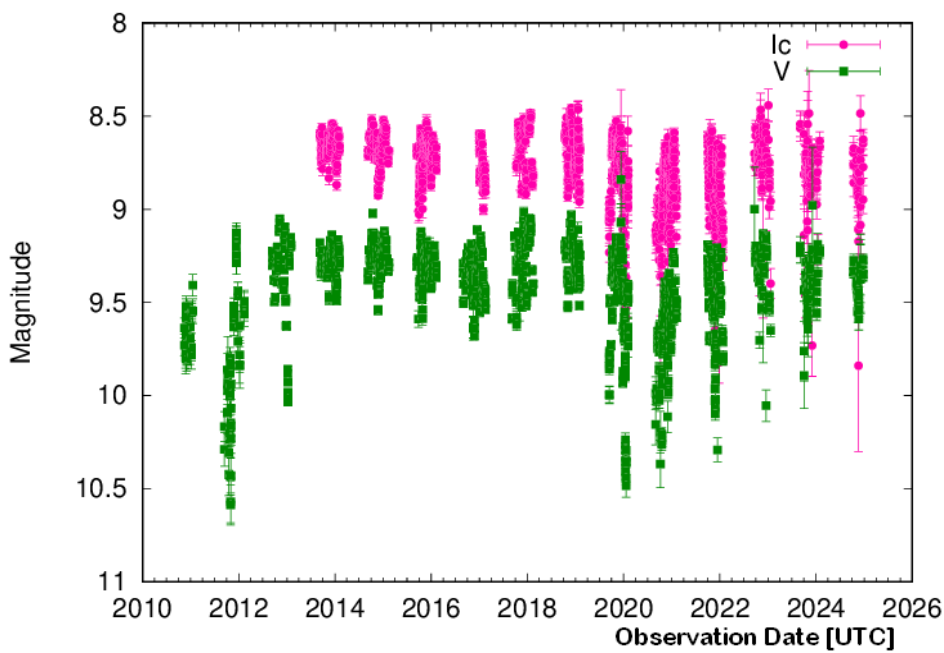

Figure 2: Ic and V brightness of XY Per in the KWS survey between 2011 and 2025

A period analysis of the KWS V data using the Lomb-Scargle GLS method in Peranso [15] reveals a prominent peak with a period of approximately 3750 days (~10 years), which corresponds to the spacing of the deeper minima. The second and third harmonics of this signal are also visible, supporting the robustness of the detected periodicity. The forest of peaks in the range of several hundred days reflects the approximate timescales of the irregular variability.

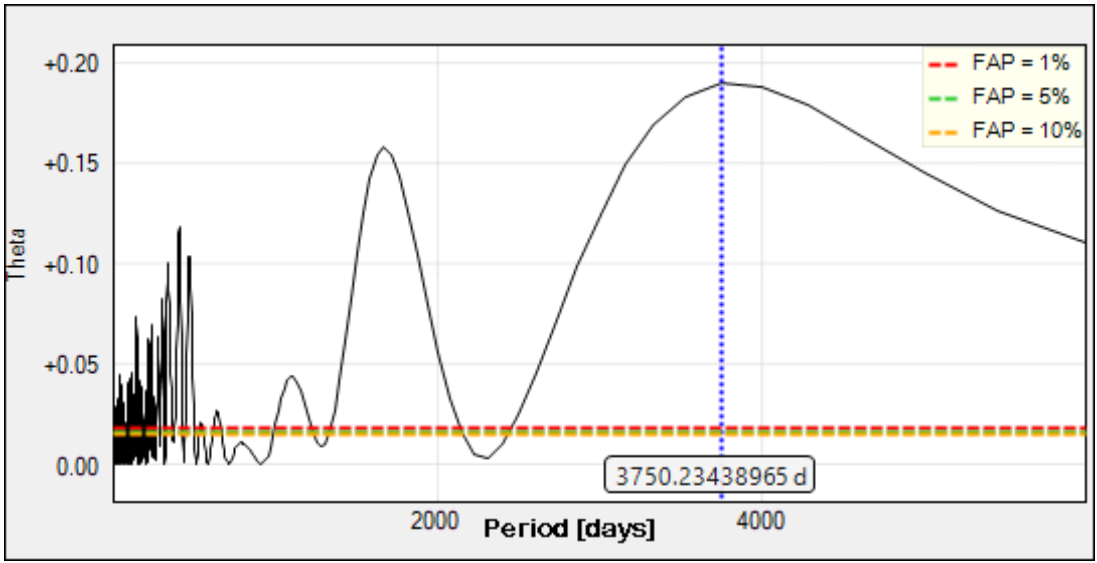

Figure 3: Peranso GLS Periodogram of the KWS V data between 2011 und 2025

To further investigate the variability of the reflection nebula, which was initially suspected by us (H.S.) based on individual DSS images, we downloaded ZTF images via the IRSA server. We created an animation composed of approximately 500 individual frames from the Zwicky Transient Facility (ZTF), covering the period from 2018 to 2024. Please note that the time intervals between individual images are not uniform. The animation, which is accessible alongside the online version of this paper reveals apparent wave-like of illumination propagating through the surrounding nebula at certain times.

Similar to XY Per, the reflection nebula also exhibits more active and less active phases. The following two pages (Figures 4-7) show two particularly active phases of the nebula and the central star during the periods August 28 – September 9, 2019, and January 15 – February 1, 2020. For comparison, the ASAS-SN and KWS data of the central star for the respective time periods are also provided.

In both events, the variability of the nebula began with a significant change in brightness of XY Per over several weeks—approximately a 0.5 mag decrease in the first case, and about a 1.5 mag increase in the second. Interestingly, although the brightness changes—as seen from Earth—occurred in opposite directions (decrease vs. increase), the response of the inner part of the reflection nebula was similar in both cases.

In both cases, a localized brightening appeared in the reflection nebula within a few days after the brightness change of the central star (red arrow, second image in the top right corner in each case). Please note, that in event 2, the exact timing of the brightness increase is not well covered due to limited observations. However, the 1.5 mag increase is still clearly visible in the second image (2020-01-23), even with saturation of the central pixels.

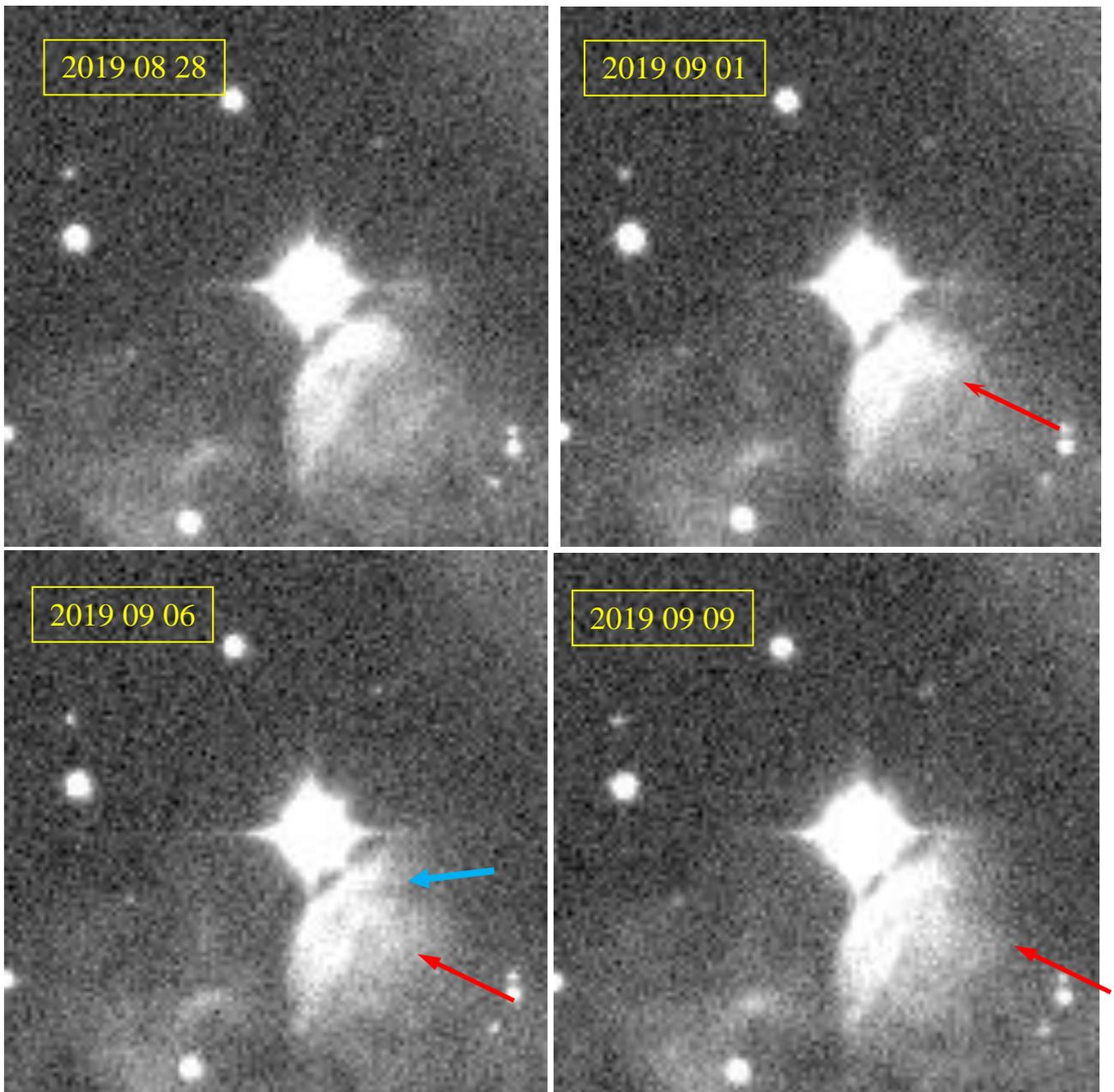

Figure 4: ZTF g images of the event between 2019-08-28 and 2019-09-09. Shown is XY Per (center) and the central region of vdB 24, (North is up, 1.8' x 1.8')

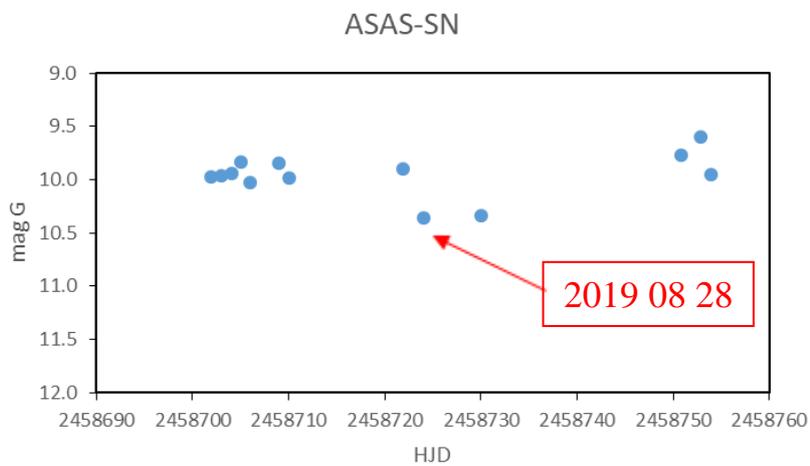

Figure 5: ASAS-SN G magnitudes around the event

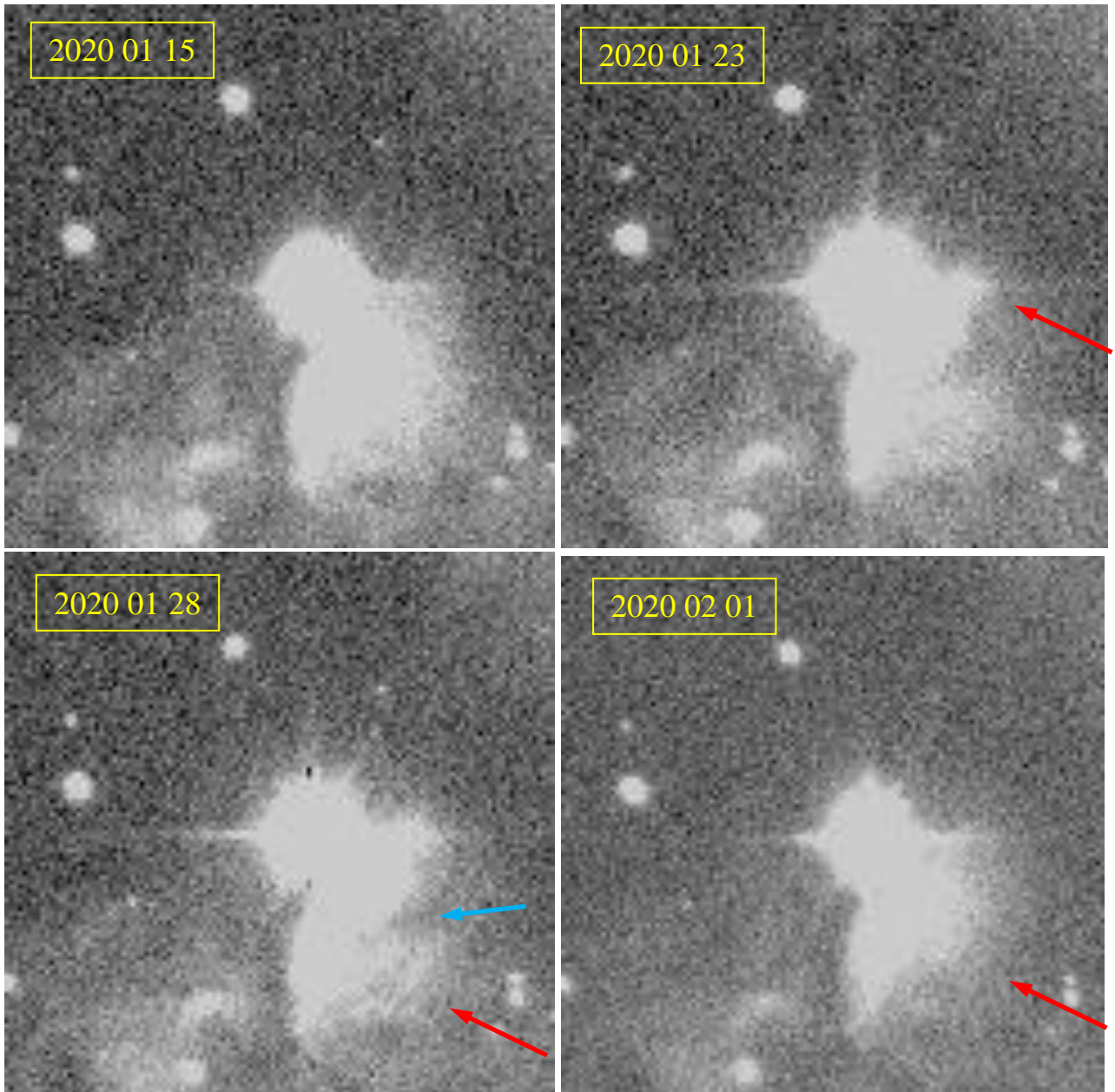

Figure 6: ZTF g images of the event between 2020-01-15 and 2020-02-01, same scale as in Figure 4

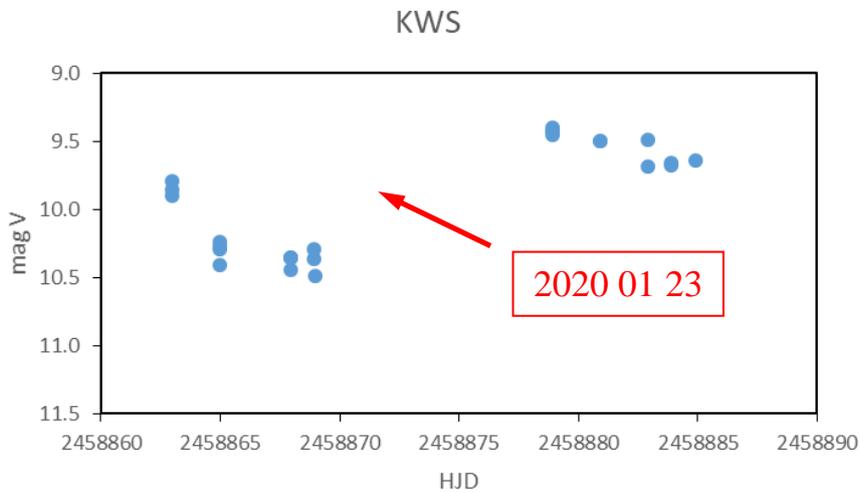

Figure 7: KWS V magnitudes around the event between 2020-01-15 and 2020-02-01.

As the event progresses over the following days, the area of localized brightening becomes larger, shifts downward, and separates from the original nebula as a gap forms (blue arrow). Eventually, in the respective fourth image, the moving brightening fades while continuing to move away, so that the nebula's original appearance is restored after about 2–3 weeks (bottom right).

An estimate based on the apparent motion of the brightening within the reflection nebula—approximately 1.5 arc seconds per day—suggests a projected propagation speed of about three to four times the speed of light, assuming a distance of around 1400 light-years (Gaia data). This seemingly superluminal motion can be explained as a geometric light echo effect, resulting from the changing illumination of surrounding dust structures by the central star. It is important to note that this is not a physical motion of matter, but rather a consequence of light travel time effects in a three-dimensional medium. This interpretation also effectively rules out alternative explanations such as the expansion of material, which would not be consistent with the observed apparent velocities.

## 4   Conclusion

XY Per and the surrounding reflection nebula VdB 24 represent a highly interesting system of an active young binary with a Herbig Ae/Be primary. This leads to position-dependent brightness changes in the surrounding reflection nebula over the course of just a few days or weeks. This object thus joins the rare class of variable nebulae, similar to Hubble's Variable Nebula. Further detailed investigations, which could also help clarify the three-dimensional structure of the nebula, are encouraged.

**Acknowledgements**

This research has utilized the SIMBAD/VIZIER database and Aladin, operated at CDS, Strasbourg, France, the International Variable Star Index (VSX) database, operated at AAVSO, Cambridge, Massachusetts, USA, the NASA/IPAC Infrared Science Archive and the SAO/NASA Astrophysics Data System, USA and the DSS2 Digitized Sky Surveys (STScI).